# Machine Learning-assisted Dynamics-Constrained Day-Ahead Energy Scheduling

Mingjian Tuo [a], Xingpeng Li [b], Pascal Van Hentenryck [c]

[a] Hubei Key Laboratory of Energy Storage and Power Battery, Hubei University of Automotive Technology, Shiyan, Hubei, 442000, CN.

[b] Department of Electrical and Computer Engineering, University of Houston, Houston, TX, 77204, USA.

[c] H. Milton Stewart School of Industrial and Systems Engineering, Georgia Institute of Technology, Atlanta, GA, 30332, USA.

---

**Abstract**

The rapid expansion of inverter-based resources, such as wind and solar power plants, will significantly diminish the presence of conventional synchronous generators in future power grids with rich renewable energy sources. This transition introduces increased complexity and reduces dynamic stability in system operation and control, with low inertia being a widely recognized challenge. However, the literature has not thoroughly explored grid dynamic performance associated with energy scheduling solutions that traditionally only consider grid steady-state constraints. This paper will bridge the gap by enforcing grid dynamic constraints when conducting optimal energy scheduling; particularly, this paper explores locational post-contingency rate of change of frequency (RoCoF) requirements to accommodate substantial inertia reductions. This paper introduces a machine learning-assisted RoCoF-constrained unit commitment (ML-RCUC) model designed to ensure RoCoF stability after the most severe generator outage while maintaining operational efficiency. A graph-informed NN (GINN)-based RoCoF predictor is first trained on a high-fidelity simulation dataset to track the highest locational

RoCoF, which is then reformulated as mixed-integer linear programming constraints that are integrated into the unit commitment model. Case studies, by solving the optimization problem ML-RCUC and validating its solutions with time-domain simulations, demonstrate that the proposed method can ensure locational RoCoF stability with minimum conservativeness.

## I. INTRODUCTION

The integration of converter-interfaced renewable energy sources (RES), particularly wind and solar power plants, has become a critical pathway for achieving electricity decarbonization objectives [1]. However, this energy transition introduces fundamental technical challenges. The displacement of conventional synchronous generation capacity progressively reduces system rotational inertia, thereby compromising power systems' ability to maintain frequency stability and secure operations during contingencies [2]. Synchronous generators traditionally provide essential grid-stabilizing functions through their inherent rotating inertia, which inherently limits frequency excursion magnitudes and suppresses excessive rate-of-change-of-frequency (RoCoF) following major disturbances [3]. As power systems evolve with reduced kinetic energy reserves from rotating masses, their vulnerability to frequency instability escalates significantly under generation-load imbalances [4]. This operational risk has manifested practically in Ireland's power system, where insufficient system inertia necessitated wind generation curtailment to preserve grid stability [5].

The accelerating integration of RES has intensified power system operators' concerns regarding frequency stability preservation. This operational paradigm shift has driven the development of frequency-constrained unit commitment (FCUC) frameworks, where dynamic RoCoF limits are incorporated into security-constrained unit commitment (SCUC) formulations - a methodology termed RoCoF-constrained unit commitment (RCUC) [6]. Such advanced scheduling models enable simultaneous determination of optimal generation portfolios and minimum synchronous inertia requirements while ensuring transient frequency security.

EirGrid has enforced a 23 GWs minimum system inertia threshold through synchronous inertial response constraints in Ireland [7]. Similarly, the Swedish grid operator ordered a 100 MW output reduction from a nuclear power plant to mitigate frequency stability risks [8]. Current research predominantly adopts system equivalent model-based RoCoF-constrained SCUC as the benchmark framework [9]-[10], which integrates frequency stability constraints derived from uniform frequency dynamics. Furthermore, [11] extended the uniform frequency response model to include converter-based control and integrated RoCoF constraints into SCUC formulations. Yang's research proposes a data-driven distributionally robust chance-constrained approach to optimize the SCUC problem while limiting the risk of frequency constraint violations [12]. Additionally, [13] investigates a mixed analytical-numerical approach based on multi-regions, combining the evolution of the center of inertia with inter-area oscillations. However, prevailing FCUC methodologies relying on COI-based frequency analysis exhibit critical limitations in modern power systems. The fundamental assumption of homogeneous inertia distribution inherent in COI formulations introduces significant modeling errors, particularly in large-scale systems with geographically dispersed and technologically diverse generation assets.

Reference [14] establishes a locational RoCoF-constrained SCUC (LRC-SCUC) framework that explicitly accounts for geographical variance in inertia distribution and network connectivity impacts on nodal frequency dynamics. The methodology formulates location-specific RoCoF stability metrics through analytical derivation, serving as a principal benchmark model. Nevertheless, simulation analyses reveal critical limita-tions in such physics-driven approach: When handling power systems' spatially distributed frequency characteristics, the inherent model simplifications fail to capture higher-order dynamics and nonlinear response patterns, resulting in approximation errors that produce either operationally infeasible or excessively conservative solutions. Overall, due to approximations, physics-driven approaches, using system equivalent model [9]-[10] and their extensions with converter control [11] or nodal frequency expression [14], may lead to large modeling errors causing either frequency security violations or overly-conservativeness.

Recent advancements attempt to overcome these limitations using data-driven approaches. Study in [15] highlights the neural networks' ability in solving unit commitment and economic dispatch problems of power systems. A neural network-assisted FCUC framework is presented to reduce both total operating cost and committed SG capacity [16]. However, the feedforward structure of conventional NNs lacks explicit mechanisms to process spatial correlations in multi-node power systems. Consequently, the model shows limited capability in tracking location-dependent frequency responses across geographically dispersed nodes - a critical shortcoming given the topological nature of modern grids. Both LRC-SCUC and NN-RCUC are implemented as benchmark models in this work to comparatively assess their spatial modeling capacities.

To address the above-identified gaps, this paper implements a novel convolutional neural network-based RoCoF-constrained unit commitment (ML-RCUC) model. A distinctive feature of this model is its ability to ensure grid frequency stability, with the proposed embedded graph-informed NN (GINN)-based RoCoF predictor, under the most severe contingency event. The proposed GINN-based predictor is trained using system operation data, capturing geographical discrepancies and locational frequency dynamics resulting from the non-uniform distribution of inertia. GINN is selected as the learner predicting RoCoF since it has the capability to capture spatial correlations. The major contributions of this work are summarized below:

- First, we present a dynamic model of a power system that realistically includes multiple generators located at the same nodes, capturing the heterogeneous responses of each node. Rather than generating random grid operation data which may be unrealistic, we utilize a SCUC model-based approach to efficiently generate reasonable data samples across a wide range of operating conditions while eliminating samples with post-contingency instability issues. Results of time-domain simulations on the generated cases are obtained for NN training.
- Secondly, to capture the difference in locational RoCoF responses which is often ignored, we establish locational post-contingency RoCoF requirements as constraints by employing a GINN-based RoCoF predictor that can process spatial information. This predictor effectively monitors locational RoCoF, even under post-contingency conditions where frequency oscillations are significant.
- Traditional RCUC methods may not capture the highest locational RoCoF on vulnerable node during oscillations. Our proposed GINN-based RoCoF predictor tracks the highest RoCoF value within the period following a large contingency, thereby ensuring frequency stability.
- The proposed ML-RCUC model, integrating the locational ML-based RoCoF predictor with piecewise linearization-based reformulation, can effectively ensure the day-ahead energy scheduling solutions are frequency-stable with least cost for all different locations.

## II. POWER SYSTEM FREQUENCY DYNAMICS

Power system frequency serves as the fundamental stability indicator. As defined in (1), aggregate system inertia represents the cumulative kinetic energy stored in synchronously rotating generator rotors [17].

$$E_{sys} = \sum_{i=1}^{N} 2H_i S_{B_i} \tag{1}$$

where $S_{B_i}$ is the generator rated power in MVA and $H_i$ denotes the inertia constant of the generator which is usually provided by the generator manufacturer.

The electromechanical dynamics governing power-frequency interactions during system disturbances are fundamentally characterized by the swing equation, as mathematically formalized in (2).

$$P_m - \Delta P_e = M\frac{\partial \Delta\omega}{\partial t} + D\Delta\omega \tag{2}$$

where $\Delta P_m$ is the total change in mechanical power and $\Delta P_e$ is the total change in electric power of the power system. $\partial\Delta\omega/\partial t$ is commonly known as RoCoF. $M$=2$H$ represents the normalized inertia constant, while $D$ denotes the damping constant [18].

The constitutive parameters collectively determine the transient frequency trajectory, enabling precise prediction of system response characteristics and informed implementation of stability control strategies. The assumption of system-wide uniformity in frequency metrics introduces critical operational vulnerabilities [13], by neglecting the geographical variance in nodal frequency dynamics across power networks. Such oversimplification compromises stability margins through inadequate representation of location-dependent oscillatory patterns. This critical oversight necessitates the development of enhanced modeling frameworks where network structural attributes and dynamic device characteristics are explicitly incorporated through nodal swing equation formulations. By establishing nodal swing equations that account for individual bus dynamics, these models fundamentally capture inter-machine oscillation patterns and localized frequency transients - essential characteristics for accurate stability assessment,

$$m_i\ddot{\theta}_i + d_i\dot{\theta}_i = P_{in,i} - \sum_{j=1}^{n} b_{i,j}\sin(\theta_i - \theta_j) \tag{3}$$

where $m_i$ and $d_i$ denote the inertia coefficient and damping ratio for node $i$ respectively, while $P_{in,i}$ denotes the power input. A network-reduced model with $N_g$ generator buses can be obtained by eliminating passive load buses via Kron reduction [19]. By focusing on network connectivity's impact on the system nodal dynamics, the phase angle $\theta$ of generator buses can be expressed by the following dynamic equation [20],

$$M\ddot{\theta}+D\dot{\theta}=P-L\theta \tag{4}$$

where $M = \operatorname{diag}(\{m_i\})$, $D = \operatorname{diag}(\{d_i\})$; for the Laplacian matrix $L$, its off-diagonal elements are $l_{ij}=-b_{ij}V_i^{(0)}V_j^{(0)}$, and diagonals are $l_{ij}=\sum_{j=1,j\neq i}^{n} b_{ij}V_i^{(0)}V_j^{(0)}$. Following disturbance $\Delta P$, the system response can be decomposed into the dynamic evolution of the frequency deviation $\delta\omega(t)$. We first expand the frequency deviation $\delta\omega(t)$ and angle deviation $\delta\theta(t)$ in terms of eigenmodes. Then solve this second-order differential equation via Laplace transforms or homogeneous/particular solutions, the RoCoF value at bus $i$ can then be derived,

$$f_{rcf,i}(t_0)=\frac{\Delta P e^{-\frac{\gamma t}{2}}}{2\pi m}\sum_{\alpha=1}^{N_g}\frac{\beta_{\alpha i}\beta_{\alpha b}}{\sqrt{\frac{\lambda_\alpha}{m}-\frac{\gamma^2}{4}}\Delta t}\left[e^{-\frac{\gamma\Delta t}{2}}\sin\left(\sqrt{\frac{\lambda_\alpha}{m}-\frac{\gamma^2}{4}}\left(t_0+\Delta t\right)\right)-\sin\left(\sqrt{\frac{\lambda_\alpha}{m}-\frac{\gamma^2}{4}}t_0\right)\right] \tag{5}$$

where $\lambda_\alpha$ is the eigenvalue, $\beta_{\alpha i}$ is the eigenvector; and $m$ denotes average inertia distribution on generator buses; and bus $b$ is where disturbance occurs; $\Delta t$ is the monitoring window, and $t_0$ is the measuring time. $N_g$ is the number of generator buses in the reduced model. The ratio of damping coefficient to inertia coefficient is $\gamma = d_i/m_i$ [4]. The overview of the working pipeline could also be found in Fig. 1.

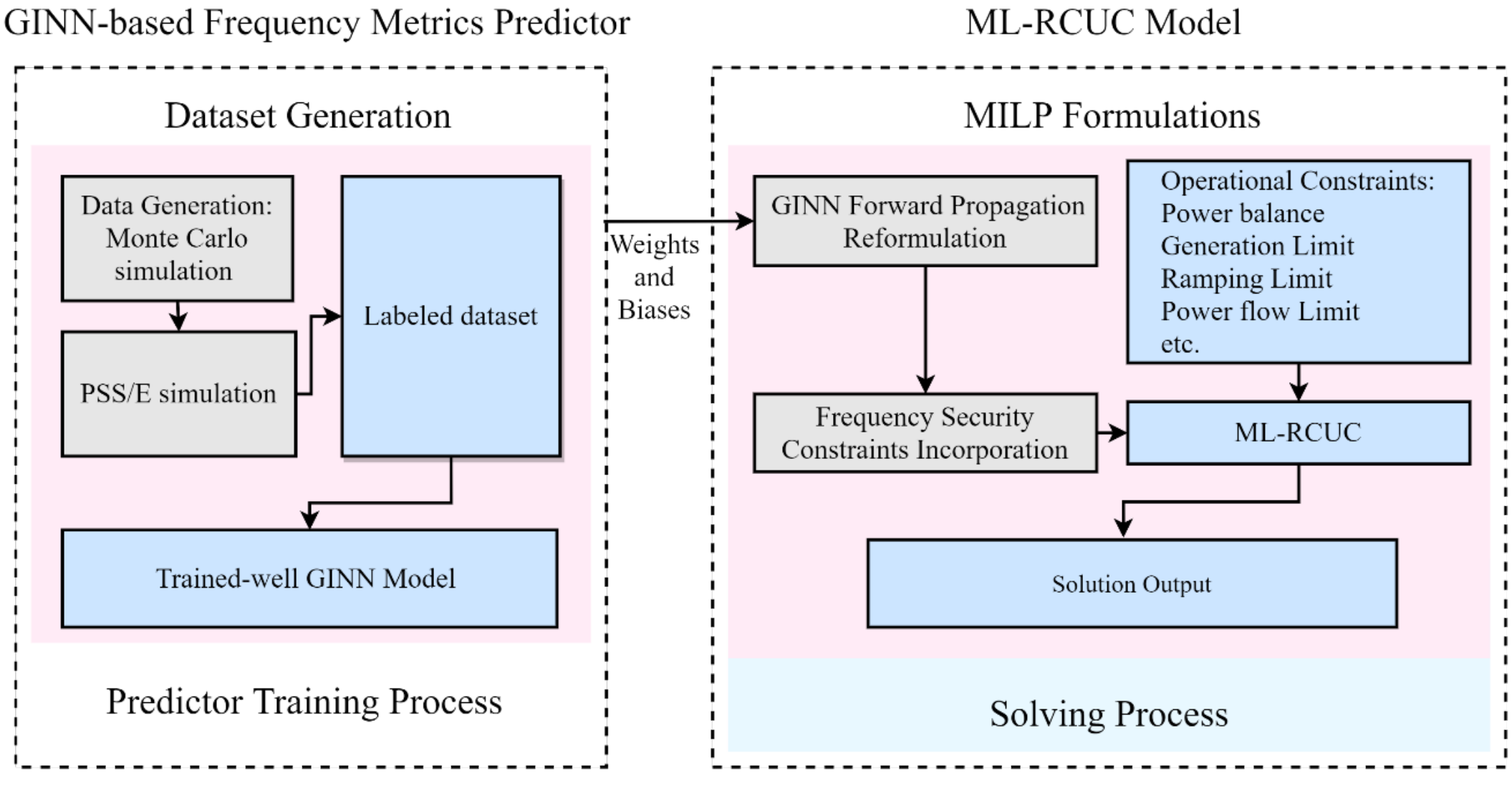

**Fig. 1.** Overview of the working pipeline.

## III. FREQUENCY STABILITY AND OSCILLATIONS

In a system with $N$ generators, the primary objective of the standard SCUC model is to minimize the total operating cost while ensuring compliance with critical technical constraints. As mathematically formalized below, the constraints typically include generator capacities, ramp rates, minimum up and down times, power balance, and transmission limits. Through optimal scheduling of generator dispatch states and output levels, the SCUC model guarantees secure and economically efficient grid operation across both normal and contingency scenarios. A generic SCUC model is as follows,

$$\begin{aligned} &\min \Gamma(s_t, u_t) \\ &\text{s.t. } \mathrm{F}(s_t, u_t, d_t, r_t) = 0, \mathrm{X}(s_t, u_t, d_t, r_t) \le 0, \forall t \end{aligned} \tag{6}$$

where $\mathrm{F}$ and $\mathrm{X}$ represent the equality and inequality constraints respectively; $s_t$ denotes the system states, and $u_t$ is the generation dispatch at period $t$; $d_t$ and $r_t$ denote the load profile and renewable forecast, respectively. It is assumed a potential disturbance $\varpi_t$ occurs in period $t$.

Traditional SCUC (T-SCUC) model only considers steady-state constraints. This model serves as a benchmark model in this paper.

$$\min_{\varphi} \sum_{g \in G} \sum_{t \in T} (c_g P_{g,t} + c_g^{NL} u_{g,t} + c_g^{SU} v_{g,t} + c_g^{RE} r_{g,t}) \tag{7a}$$

$$\sum_{g \in n} P_{g,t} + \sum_{k \in K^-(n)} P_{k,t} - \sum_{k \in K^+(n)} P_{k,t} - D_{n,t} + E_{n,t} = 0, \forall n, t \tag{7b}$$

$$P_{k,t} - b_k \left( \theta_{n,t} - \theta_{m,t} \right) = 0, \forall k, t \tag{7c}$$

$$-P_k^{\max} \le P_{k,t} \le P_k^{\max}, \forall k, t \tag{7d}$$

$$P_g^{\min} u_{g,t} \le P_{g,t}, \forall g, t \tag{7e}$$

$$P_{g,t} + r_{g,t} \le u_{g,t} P_g^{\max}, \forall g, t \tag{7f}$$

$$0 \le r_{g,t} \le R_g^{re} u_{g,t}, \forall g, t \tag{7g}$$

$$\sum_{j \in G} r_{j,t} \ge P_{g,t} + r_{g,t}, \forall g, t \tag{7h}$$

$$P_{g,t} - P_{g,t-1} \le R_g^{hr}, \forall g, t \tag{7i}$$

$$P_{g,t-1} - P_{g,t} \le R_g^{hr}, \forall g, t \tag{7j}$$

$$v_{g,t} \ge u_{g,t} - u_{g,t-1}, \forall g, t \tag{7k}$$

$$v_{g,t+1} \le 1 - u_{g,t}, \forall g, t \le nT - 1 \tag{7l}$$

$$v_{g,t} \le u_{g,t}, \forall g,t \tag{7m}$$

$$\sum\nolimits_{s=t-UT_g}^{t} v_{g,s} \le u_{g,t}, \forall g, t \ge UT_g \tag{7n}$$

$$\sum\nolimits_{s=t-UT_g}^{t+DT_g} v_{g,s} \le 1 - u_{g,t}, \forall g, t \ge nT - DT_g \tag{7o}$$

$$u_{g,t}, v_{g,t} \in \{0,1\}, \forall g,t \tag{7p}$$

Equation (7a) represents the objective function, with the fundamental constraints outlined in Equations (7b) to (7o). Specifically, Equation (7b) enforces nodal power balance, while Equation (7c) calculates network power flows, which are limited by the transmission capacities specified in Equation (7d). Scheduled energy production and generation reserves are constrained by unit generation capacity and ramping rates, as defined in Equations (7e) through (7j). According to Equation (7h), reserve requirements ensure adequate coverage for the loss of a single generator. The start-up and operational status of conventional units are defined as binary variables in Equations (7k) to (7m). The minimum downtime before a generator can be restarted and the minimum uptime before it can be shut down are detailed in Equations (7n) and (7o), respectively. Finally, the binary variables indicating generator start-up and commitment status are represented in Equation (7p).

While the T-SCUC model has been used in the industry for decades, additional constraints are required when enforcing system frequency stability addressing the emerging low-inertia grid challenge. Physical model-based RoCoF constraints can be derived from (3) or (5) and subsequently integrated into the standard SCUC formulation, forming enhanced models like ERC-SCUC or LRC-SCUC [14], to ensure system frequency stability. However, the assumptions made during these RoCoF expression derivation processes may introduce large approximation errors, potentially resulting in either insecure or overly conservative outcomes. The proposed data-driven approach is to address these limitations by replacing physics-based constraints with ML formulations. Frequency predictor is trained on physically plausible operational scenarios generated via a SCUC models, avoiding unrealistic samples that could bias predictions. Its graph-informed architecture explicitly captures spatial correlations in multi-node power systems, a capability lacking in conventional neural networks. This method leverages the predictive capabilities of neural networks to more accurately capture system dynamics and improve the reliability of the SCUC model as in (8),

$$\hat{h}^f\left(s_t, u_t, r_t, \varpi_t\right) \leq \varepsilon, \forall t \tag{8}$$

where $\hat{h}^f$ is the nonlinear neural network-based predictor for mapping system condition and disturbance onto the frequency stability metric that is maximal locational RoCoF; $\varepsilon$ is the predefined threshold. The loss of generation not only results in the power mismatch, but also leads to the reduction in system inertia, which further leads to the highest RoCoF value and frequency deviation; thus, $G-1$ event is considered as the worst contingency instead of sudden load increase in this study

## IV. GINN-BASED RoCoF PREDICTION

This research pioneers a machine learning-driven paradigm shift in frequency stability assessment, circumventing the inherent simplifications of conventional physics-derived RoCoF formulations. The developed data-driven framework establishes a nonlinear mapping between multidimensional operational inputs-including contingency characteristics, real-time generation dispatch patterns, and measurement node configurations-and the resultant locational RoCoF distribution across the power network. Capitalizing on GINN's capability to extract complex spatiotemporal dependencies, this approach achieves superior approximation accuracy compared to analytical methods, particularly in capturing geographically correlated frequency response dynamics. This integration enables adaptive stability boundary adjustments that account for real-time system topology and operational states, effectively bridging the fidelity gap between traditional security-constrained optimization and transient stability requirements.

*A. Power System Feature Definition*

As both the magnitude and location of contingencies impact locational frequency deviation and inertial response, generator status and dispatch values are encoded into feature vectors [21]. For a given time period *t*, the generator status vector is defined below,

$$u_t = \left[ u_{1,t}, u_{2,t}, \cdots, u_{g,t} \right], \forall g \in G, \forall t \tag{9}$$

The disturbance feature vector is defined against the loss of the largest generation, the magnitude of the contingency could be expressed as,

$$P_t^{con} = \max_{g \in G} \left( P_{1,t}, P_{2,t}, \cdots, P_{g,t} \right), \forall t \tag{10}$$

The location of the disturbance is represented by the index of the generator producing maximum power,

$$g_t^{con} = \arg \max_{g \in G} \left( P_{1,t}, P_{2,t}, \cdots, P_{g,t} \right), \forall t \tag{11}$$

The information of magnitude and location into the disturbance feature vector as,

$$\varpi_t^G = \left[ 0, \cdots, 0, \underbrace{P_t^{con}}_{g_t^{con} \text{th element}}, 0, \cdots, 0 \right], \forall t \tag{12}$$

*B. Model-based Sample Generation*

While randomized nodal power injection strategies provide broad operational scenario coverage, their unconstrained application introduces critical limitations in large-scale power system modeling. The unguided randomization process tends to generate physically implausible grid configurations, exponentially inflating dataset requirements while simultaneously degrading prediction accuracy through inclusion of non-representative operating states. More critically, such artificial scenarios frequently induce premature numerical instabilities in time-domain simulations, thereby invalidating contingency analysis and potentially obscuring genuine stability risks under credible disturbance conditions.

To address these challenges, we implemented a model-based approach for effectively generating training datasets, offering a more predictable alternative to random data generation. This methodology leverages specialized SCUC models to generate data system-atically through Monte Carlo simulations, encompassing a wide range of load conditions and RES scenarios. The primary aim is to minimize operational costs, including those related to fuel usage, idle capacities, startup procedures, and the maintenance of adequate reserves. Paper [14] provides a detailed explanation of how a piecewise linear programming method is introduced to incorporate nonlinear locational RoCoF-related constraints into a mixed integer linear programming (MILP) model.

Scenario data is then integrated into PSS/E model for dynamic simulation. First, the Pyomo-based framework generates operational parameters that define each scenario, including generator on/off statuses, dispatch levels, load profiles, renewable generation forecasts, and transmission line statuses. These data are output in standardized CSV formats with clear mappings to PSS/E's data structures. Second, we use PSS/E's API functions to directly update PSS/E's model data. This includes setting generator active power setpoints, updating load magnitudes at respective buses, and configuring transmission line statuses—all aligned with the scenario constraints enforced by Pyomo. This direct API interaction ensures that the steady-state conditions initialized in PSS/E exactly mirror the Pyomo-generated scenarios. After initializing the initial power flow, time-domain simulations are executed programmatically through PSS/E's API. This includes triggering the contingency event and specifying simulation parameters.

### *C. GINN-based RoCoF Predictor Development*

ML techniques have demonstrated significant efficacy in processing high-dimensional power system data by automatically extracting complex spatiotemporal correlations. Capitalizing on hierarchical feature learning capabilities, these data-driven models excel at identifying multi-scale operational patterns from system-wide measurements. Recent advances in neural computing architecture reveal that graph-informed frameworks can effectively capture localized dynamic interactions within power networks while preserving global stability characteristics. The extracted latent representations, encoding critical relationships between generation dispatch, network topology, and stability margins, substantially enhance decision-making fidelity in SCUC optimization through improved feature interpretability and generalization capacity [21].

The GINN model employed in this study is depicted in Fig. 2, created using the NN-SV Tools. The model comprises two primary types of layers: convolutional and fully connected layers. The convolutional layers are designed to learn feature representations of the input data. Each convolutional layer consists of multiple convolutional kernels, which compute different feature maps. Neurons within a feature map are connected to neighboring neurons in the preceding layer, establishing a receptive field for each neuron. This receptive field determines the region of influence for the neuron's input.

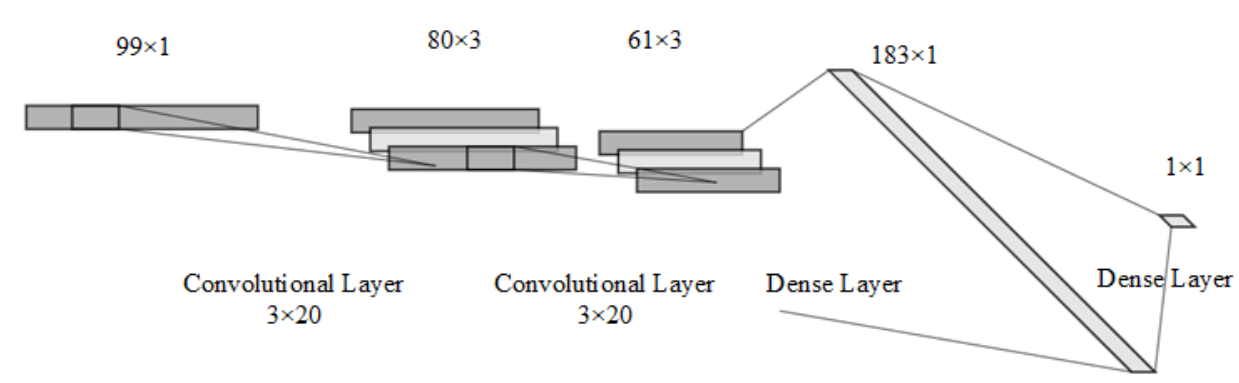

**Fig. 2.** Architecture of proposed GINN model.

The feature transformation process employs learnable parameterized filters to extract discriminative patterns from input data through sliding window operations, as mathematically defined in (13)-(14), the architecture ensures systematic pattern extraction while maintaining computational efficiency through weight reuse, with the activation functions enabling hierarchical feature learning critical for capturing multi-scale system dynamics.

$$\hat{z}_{i,j,\xi}^{o} = x_{i,j} w_{\xi}^{o} + b_{\xi}^{o}, \forall i, \forall j, \forall o, \forall \xi \tag{13}$$

$$z_{i,j,\xi}^{o} = \max\left(\hat{z}_{i,j,\xi}^{o}, 0\right), \forall i, \forall j, \forall o, \forall \xi \tag{14}$$

where $w_{\xi}^{o}$ and $b_{\xi}^{o}$ are the weight vector and bias term of $\xi$-th filter of $o$-th layer respectively, and $x_{i,j}$ is the input patch centered at location $(i, j)$ of $o$-th layer. Noted that the kernel $w_{\xi}^{o}$ that generates the feature map is shared; such mechanisms can reduce the model complexity and improve the efficiency of the model. Rectified linear unit (ReLU) is used as the activation function for introducing nonlinearities to GINN model.

Prior to dimensionality reduction, the hierarchical feature representations undergo vectorization to transform multidimensional tensors into sequential vectors compatible with feedforward NN architectures. This critical preprocessing step ensures structural compatibility between feature encoding space and subsequent decision-making layers. Denoting the flatten function as $\text{flatten}(\cdot)$, for the generated feature map $z^{o}_{i,j,\xi}$ we have,

$$z^{full} = \text{flatten}\left(z^{O_N}_{\xi}\right) \tag{15}$$

$$z^{1} = z^{full}W^{1} + b^{1} \tag{16}$$

$$\hat{z}^{q} = z^{q-1}W^{q} + b^{q} \tag{17}$$

$$z^{q} = \max\left(z^{q}, 0\right) \tag{18}$$

and

$$R_{h,s} = z_{N_L}W_{N_L+1} + b_{N_L+1} \tag{19}$$

where $O_N$ is the last convolutional layer, $W^q$ and $b^q$ represent the weight and bias for the $q$-th fully connected layer. $W_{N_L+1}$ and $b_{N_L+1}$ represent the set of weight and bias of the output layer. The training process is to minimize the total mean squared error between the predicted output and the labeled outputs of all training samples as follows,

$$\min_{\Phi} \frac{1}{N_S}\sum_{s=1}^{N_S}\left(\dot{f}_{\max} - \hat{f}_{rcf}\right)^{2} \tag{20}$$

where $\Phi = \left\{w^{o}_{\xi}, b^{o}_{\xi}, W^{q}, b^{q}, W^{rcf}_{N_L+1}, b^{rcf}_{N_L+1}\right\}$ represents the set of optimization variables.

## V. MACHINE LEARNING ASSISTED ROCOF-CONSTRAINED SCUC

The proposed ML-RCUC model is explained in this section. The unique advantage of this enhanced ML-RCUC model is that it includes the GINN-based RoCoF Predictor to establish frequency performance constraints such that the proposed RCUC model can ensure system frequency stability. The detailed formulation is presented below.

### A. Feature Encoding

Since $\varpi^{G}_{t}$ includes the max operator, it cannot be directly utilized in the encoding formulation. To address this, we introduce supplementary variables to indicate whether a generator is the largest active power generation source during the scheduling period. The reformulated equations are expressed as follows:

$$P^{G}_{\chi,t} - P^{G}_{g,t} \le A\left(1 - v^{G}_{g,t}\right), \forall g, t \tag{21}$$

$$\sum_{g\in G} v_{g,t}^{G} = 1, \forall t \tag{22}$$

where $A$ is a big number. Equation (21) enforces $v_{g,t}^{G}$ to be zero if the dispatched power of any generator $\chi$ is larger than interested generator $g$ at period $t$. Equation (22) ensures that there would be only one largest generator being considered as the potential largest contingency. Equation (21) together with (22) could enforce generator $g$ has the largest output power and $v_{g,t}^{G}$ would be set as 1. To further encode the magnitude and spatial information of disturbance into feature vector, variable $\rho_{\chi,t}^{G}$ is introduced, and the value of $\rho_{\chi,t}^{G}$ equals to the largest generation $P_{g,t}^{G}$, the constraints can be expressed as follows,

$$\rho_{\chi,t}^{G} - P_{g,t}^{G} \geq -A\left(1 - v_{g,t}^{G}\right), \forall g,t \tag{23}$$

$$\rho_{\chi,t}^{G} - P_{g,t}^{G} \leq A\left(1 - v_{g,t}^{G}\right), \forall g,t \tag{24}$$

$$0 \leq \rho_{\chi,t}^{G} \leq A v_{g,t}^{G}, \forall g,t \tag{25}$$

Thus, the overall feature vector $x_t$ can be then defined as follows,

$$x_t = [u_{1,t}, \cdots, u_{g,t}, \rho_{1,t}^{G}, \cdots, \rho_{g,t}^{G}, P_{1,t}, \cdots, P_{g,t}], \forall g,t \tag{26}$$

*B. GINN Constraints Linearization*

Since ReLU activation functions are nonlinear, to include the NN into the MILP, binary variables $\alpha_{i,j,\xi,t}^{o}$ and $\alpha_{l,t}^{q}$ are introduced which represent the activation status of the ReLU within NN model. Considering $A$ is a big number that is larger than the absolute value of all $\hat{z}_{i,j,\xi,t}^{o}$ and $\hat{z}_{l,t}^{q}$, when preactivated value $\hat{z}_{i,j,\xi,t}^{o}$ or $\hat{z}_{l,t}^{q}$ is larger than zero, constraints (27b) – (27e) and (27h) – (27l) will force binary variables $\alpha_{i,j,\xi,t}^{o}$ or $\alpha_{l,t}^{q}$ to one, and the activated value will be equal to pre-activated value. When $\hat{z}_{i,j,\xi,t}^{o}$ or $\hat{z}_{l,t}^{q}$ is less than or equal to zero, constraints will force binary variable $\alpha_{i,j,\xi,t}^{o}$ or $\alpha_{l,t}^{q}$ to zero. Subsequently, the activated value will be set zero.

$$\hat{z}_{i,j,\xi,t}^{o} = x_{i,j,t} W_{\xi}^{o} + b_{\xi}^{o}, \forall i,j,t,o,\xi \tag{27a}$$

$$z_{i,j,\xi,t}^{o} \leq \hat{z}_{i,j,\xi,t}^{o} + A\left(1 - \alpha_{i,j,\xi,t}^{o}\right), \forall i,j,t,o,\xi \tag{27b}$$

$$z_{i,j,\xi,t}^{o} \geq \hat{z}_{i,j,\xi,t}^{o}, \forall i,j,t,o,\xi \tag{27c}$$

$$z_{i,j,\xi,t}^{o} \leq A\alpha_{i,j,\xi,t}^{o}, \forall i,j,t,o,\xi \tag{27d}$$

$$z^{o}_{i,j,\xi,t} \geq 0, \forall i,j,t,o,\xi \tag{27e}$$

$$z^{full}_{t} = \text{flatten}\left(z^{O_N}_{t}\right), \forall t \tag{27f}$$

$$z^{1}_{l,t} = z^{full}_{t} W^{1}_{l} + b^{1}_{l}, \forall l,t \tag{27g}$$

$$\hat{z}^{q}_{l,t} = z^{q-1}_{l,t} W^{q}_{l} + b^{q}_{l}, \forall q,l,t \tag{27h}$$

$$z^{q}_{l,t} \leq \hat{z}^{q}_{l,t} + A\left(1-\alpha^{q}_{l,t}\right), \forall q,l,t \tag{27i}$$

$$z^{q}_{l,t} \geq \hat{z}^{q}_{l,t}, \forall q,l,t \tag{27j}$$

$$z^{q}_{l,t} \leq A\alpha^{q}_{l,t}, \forall q,l,t \tag{27k}$$

$$z^{q}_{l,t} \geq 0, \forall q,l,t \tag{27l}$$

$$\alpha^{o}_{i,j,\xi,t}, \alpha^{q}_{l,t} \in \{0,1\}, \forall i,j,o,q,\zeta,l,t \tag{27m}$$

## VI. RESULTS ANALYSIS

A case study on the IEEE 24-bus system [22] is presented to demonstrate the effectiveness of the proposed methods. The test system includes 24 buses, 33 generators, 38 transmission lines, and incorporates wind and solar power as renewable resources. The model-based data generation is implemented through Pyomo-based framework in Python [23], systematically producing grid operating conditions and generation dispatch profiles adhering to N-1 security criteria. For stability assessment, time-domain simulations are executed in PSS/E [24], with frequency responses serving as stability labels for machine learning training. Renewable generation profiles, spanning 24-hour operational cycles with ±30% power fluctuations, are illustrated in Fig. 3 to reflect realistic variability patterns.

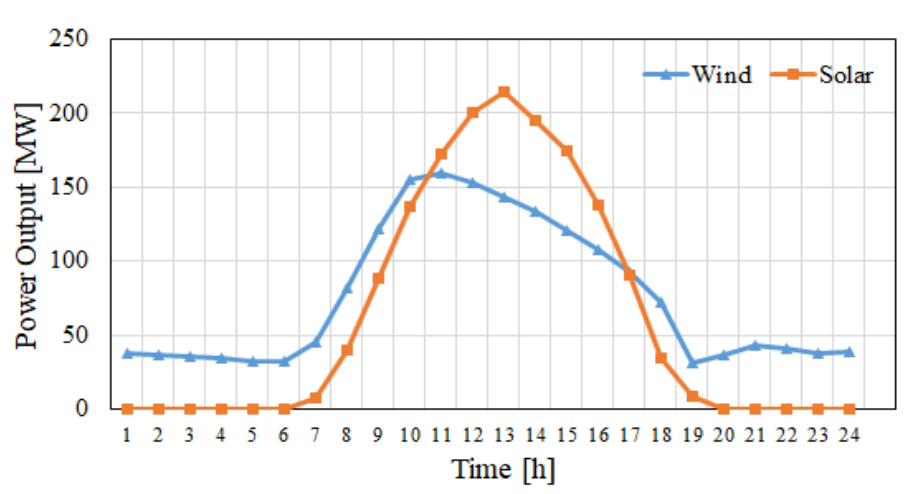

**Fig. 3.** RES generations forecasts.

For dynamic simulations during labeling, full-scale models are utilized: GENROU and GENTPJ for synchronous machines, IEEEX1 for the excitation system, IEESGO for the turbine-governor, and PSS2A for the power system stabilizer. Standard wind turbine generator models and their corresponding con-trol modules are also employed. The data creation and verification processes are executed using Pyomo and PSS/E, with the SCUC problem solved by Gurobi.

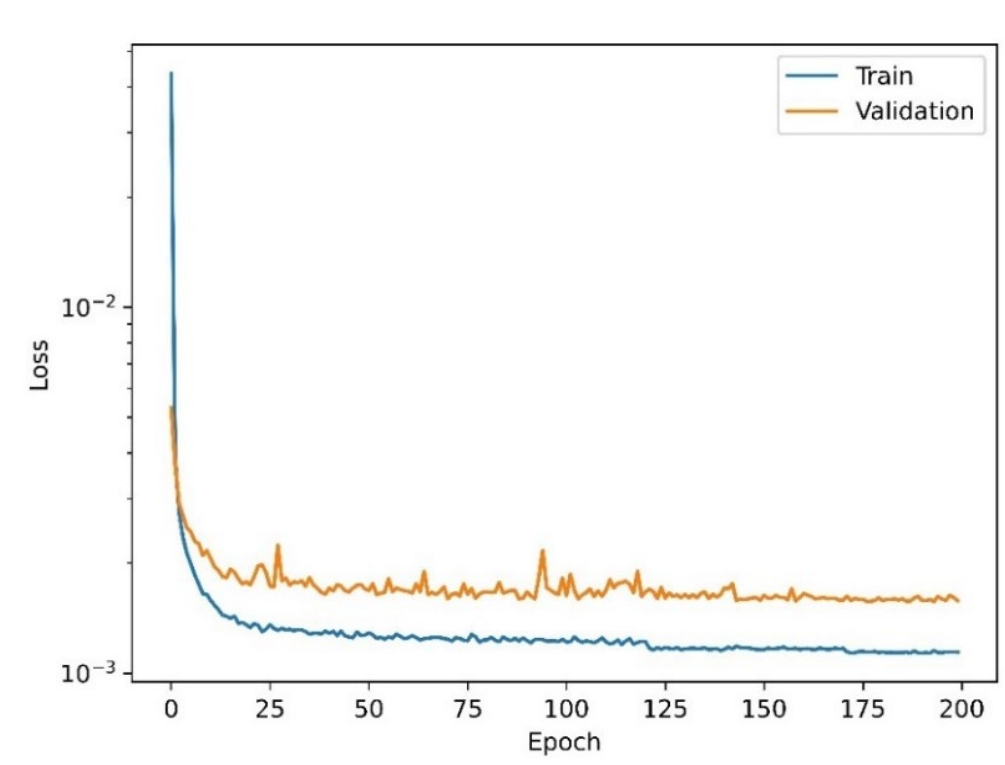

**Fig. 4.** Learning curve of GINN model.

*A. NN-based RoCoF Predictor Training*

The model-based dataset with 8,300 operational scenarios was systematically generated for training the ML-based stability predictor. Each scenario is annotated with transient security labels (0: insecure, 1: secure) derived from post-contingency RoCoF and frequency nadir analysis. To ensure operational realism and model generalizability, load and renewable generation profiles were sampled from truncated Gaussian distributions (±20% deviation from nominal values), with system constraints strictly enforced through a 0.1% optimality gap in the SCUC solver. Generator operational limits include synchronous unit reactive power capabilities and wind farm unity power factor operation [23].

Several common classification methods are first compared with the proposed GINN model using data from the IEEE 24-bus system. As shown in TABLE I, the performance of the GINN-based RoCoF predictor is evaluated against a benchmark fully-connected DNN. The validation accuracy of the GINN-based RoCoF predictor is compared with the benchmark DNN model across different error tolerances. At a 5% error tolerance, the GINN model achieves a validation accuracy of 93.53%, slightly higher than the DNN model's 92.78%. When the error tolerance increases to 10%, the GINN model's validation accuracy improves to 97.96%, compared with DNN’s 97.49%. With 15% error tolerance, the GINN model further outperforms the DNN model, achieving a validation accuracy of 99.17% versus the DNN model's 98.93%. Fig. 4 depicts the evolution of Mean Squared Error (MSE) losses on both training and validation sets throughout the training process for the proposed GINN

TABLE I. VALIDATION ACCURACY

| Error Tolerance | 5% | 10% | 15% |
|---|---|---|---|
| DNN | 92.78% | 97.49% | 98.93% |
| GINN | 93.53% | 97.96% | 99.17% |

model. These results clearly demonstrate the superior prediction accuracy of the GINN model.

Moreover, the GINN predictor was evaluated with extra metrics to assess prediction accuracy: (1) median absolute error (MED-E), (2) mean absolute error (MEA-E), and (3) R2 score. TABLE II shows that the GINN-based RoCoF predictor achieves lower MSE and MEA-E values, indicating superior performance in processing power system data with embedded graphical information.

TABLE II. MODEL COMPARISONS

| Model | MED-E | MEA-E | $R^2$ |
|---|---|---|---|
| DNN | 0.0072 | 0.0039 | 0.9828 |
| GINN | 0.0055 | 0.0021 | 0.9893 |

*B. SCUC Simulation Results Analysis*

In addition to the traditional SCUC model, we also compare the proposed ML-based RCUC model with two physics-based SCUC models enforcing frequency performance [14]: (i) system equivalent model-based RoCoF-constrained SCUC, or ERC-SCUC, and (ii) location-based RoCoF-constrained SCUC, or LRC-SCUC.

Electricity demand in the study is set to range from 1,300 MW to a peak of 1,786 MW. As shown in TABLE III, all RoCoF-constrained SCUC models effectively reduce reserve costs compared to the traditional SCUC model. Notably, the ML-RCUC model reduces reserve costs from $83,475 to $61,882, reflecting a decrease of 25.87%. However, the results in TABLE III show an increase in startup costs when RoCoF-related constraints are applied in period 19, suggesting additional costs are incurred to enhance generator flexibility.

TABLE III. COMPARISON OF DIFFERENT MODELS' COSTS [$]

| Model | Total | Operational | Startup | Reserve |
|---|---|---|---|---|
| T-SCUC | 442,978 | 336,846 | 22,657 | 83,475 |
| ERC-SCUC | 456,045 | 369,820 | 23,782 | 62,443 |
| LRC-SCUC | 456,122 | 369,084 | 23,782 | 62,256 |
| NN-RCUC | 456,178 | 370,241 | 23,782 | 62,155 |
| ML-RCUC | 456,294 | 370,630 | 23,782 | 61,882 |

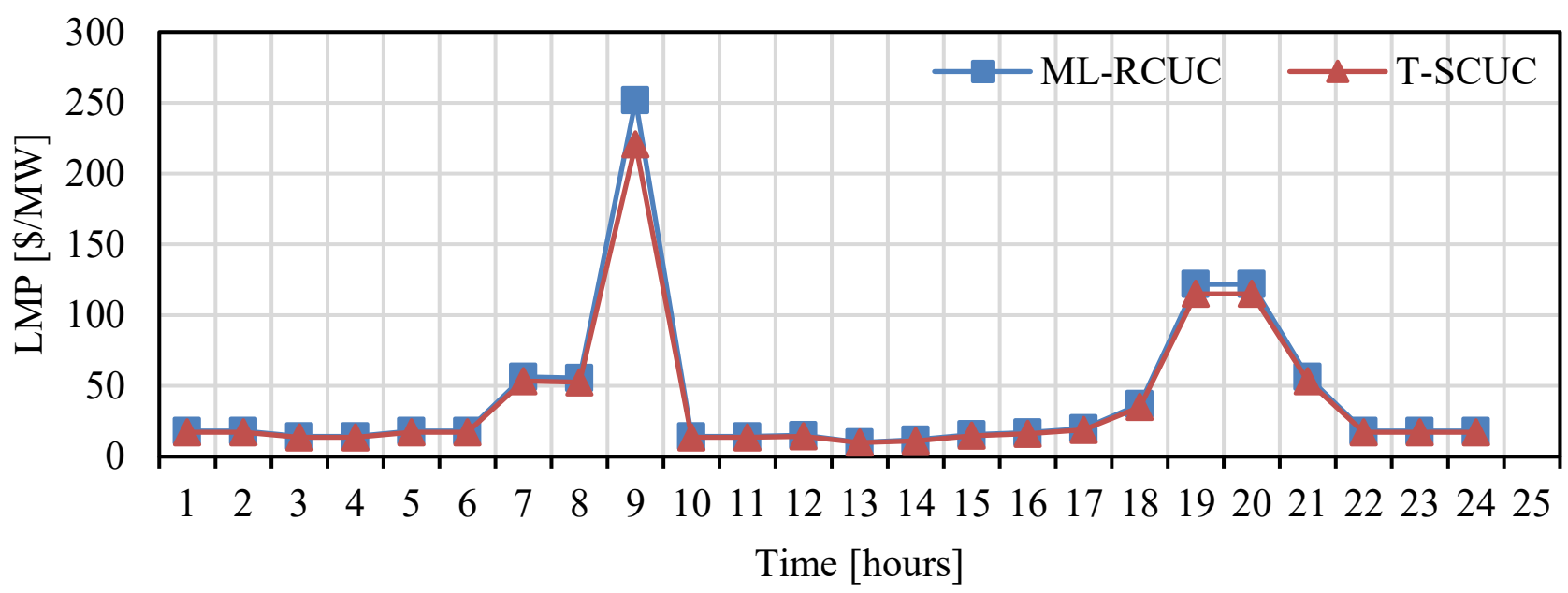

**Fig. 5.** Locational marginal price of different models.

TABLE IV Market Results (Averaged Over 24 Hrs)

| Model | Load payment [$/h] | Generator revenue [$/h] | Congestion revenue [$/h] |
|---|---|---|---|
| T-SCUC | 47,307 | 56,649 | -9,342 |
| ML-RCUC | 90,499 | 41,416 | 49,083 |

Fig. 5 compares the 24-hour average locational marginal price (LMP) of different cases. Both models start with relatively low LMPs, but around the 8th hour, the LMP of ML-RCUC begins to diverge slightly higher than that of T-SCUC. Notably, ML-RCUC exhibits an extreme price spike absent in T-SCUC, which indicates that when extra generators are committed to acquire additional inertia and enhance frequency stability in the ML-RCUC scenario, a significant price increase occurs, leading to reduced system efficiency and increased operation cost. During the evening hours, another noticeable LMP increase is observed in the ML-RCUC model, while the T-SCUC model also sees an increase but to a lesser extent. As demand likely drops toward the day's end, both models show a decrease in LMP, and their LMPs converge at similar values, suggesting reduced congestion or demand. Compared with the T-SCUC model, the ML-RCUC model demonstrates greater price volatility with significant spikes.

According to the market results in TABLE IV, the sensitivity of the ML-RCUC model to GINN-based frequency-related constraints and the integration of renewable energy sources lead to higher congestion costs. This indicates that with GINN-based constraints included, network congestion contributes more to the differences in nodal LMPs and increases load payment.

### *C. Generators On/Off Status Analysis*

The results depicted in Figs. 6 and 7 illustrate the generator on/off status for different cases. Due to the integration of renewable energy, generators in both cases are shut down from hour 10 to hour 17, in contrast to the traditional scenario without frequency-related constraints. The traditional SCUC case demonstrates a significantly lower number of committed generators compared to the proposed ML-RCUC model, indicating that additional generators are required to be online and synchronized in the system to ensure frequency stability. The results also reveal that generators in the northern regions are more frequently turned on compared to those in the southern regions, highlighting the regional characteristics of inertia distribution and frequency dynamics.

The incorporation of RES significantly influences the operational strategy, necessitating the activation of additional generators to maintain stability. This is particularly evident in the proposed GINN-embedded ML-RCUC model, where the number of operational online generators is increased to compensate for the variability and lower inertia associated with RES. The difference in the fleet of online generators between northern and southern regions underscores the importance of considering regional inertia characteristics and frequency dynamics in unit commitment strategies.

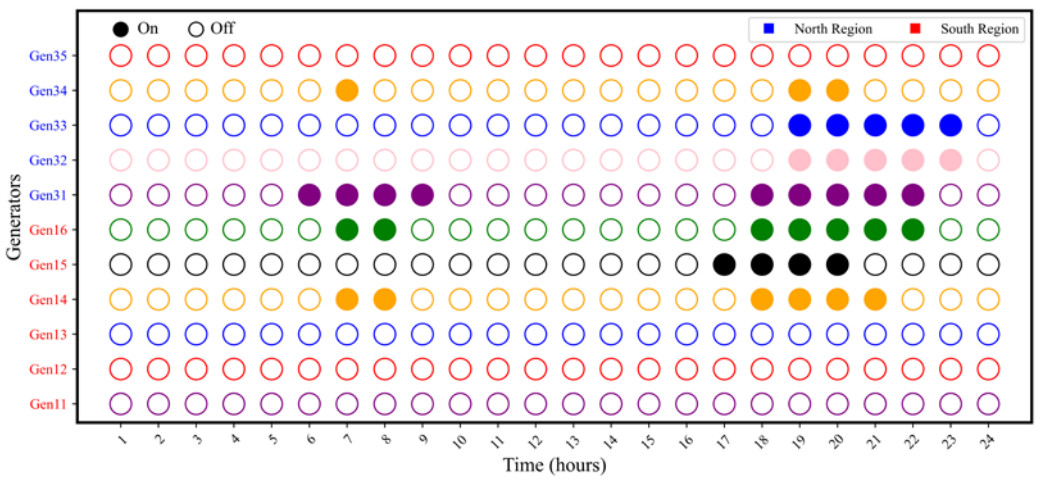

**Fig. 6.** Generator on/off status of T-SCUC.

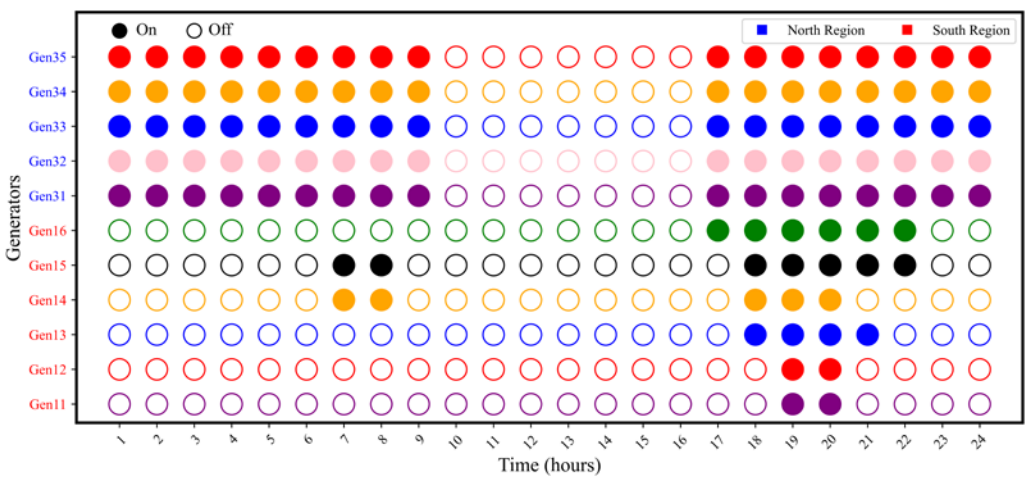

**Fig. 7.** Generator on/off status of ML-RCUC.

### *D. Time Domain Simulation Results Analysis*

We assume the worst-case contingency occurs in hour 19, where the generator with the highest output power is tripped. The system uniform RoCoF responses for various scheduling cases are illustrated in Fig. 7. Fig. 7(a) displays the uniform RoCoF response of the ERC-SCUC model. With system equivalent RoCoF constraints incorporated, the maximum RoCoF absolute value for the ERC-SCUC is strictly limited to 0.5 Hz/s, thereby meeting the RoCoF constraint. Figs. 7(b), (c), and (d) show that the system uniform RoCoF does not exceed the threshold for both the LRC-SCUC and DNN-RCUC models. Notably, the proposed ML-RCUC model exhibits the lowest RoCoF value.

The locational RoCoF dynamics for all models are plotted in Fig. 9. Combining the insights from Figs. 8(a) and 9(a), we observe that although the system uniform RoCoF remains within the threshold with the equivalent RoCoF constraint applied, the locational RoCoF at several generator buses exceeds its threshold due to oscillations. Consequently, the ERC-SCUC is deemed insecure, as the dispatched condition cannot withstand the trip of the largest generator, potentially leading to a cascaded generator contingency. Fig. 9(b) shows that while the LRC-SCUC schedule mitigates the highest RoCoF, it still exceeds the threshold due to approximation errors. Data-driven methods, however, demonstrate improved RoCoF dynamics. Fig. 9(c) reveals that the DNN-RCUC can reduce the highest RoCoF to -0.53 Hz/s, which slightly breaches the threshold. In contrast, as shown in Fig. 9(d), the proposed ML-RCUC method achieves the highest RoCoF of -0.52 Hz/s, outperforming all other SCUC schedules.

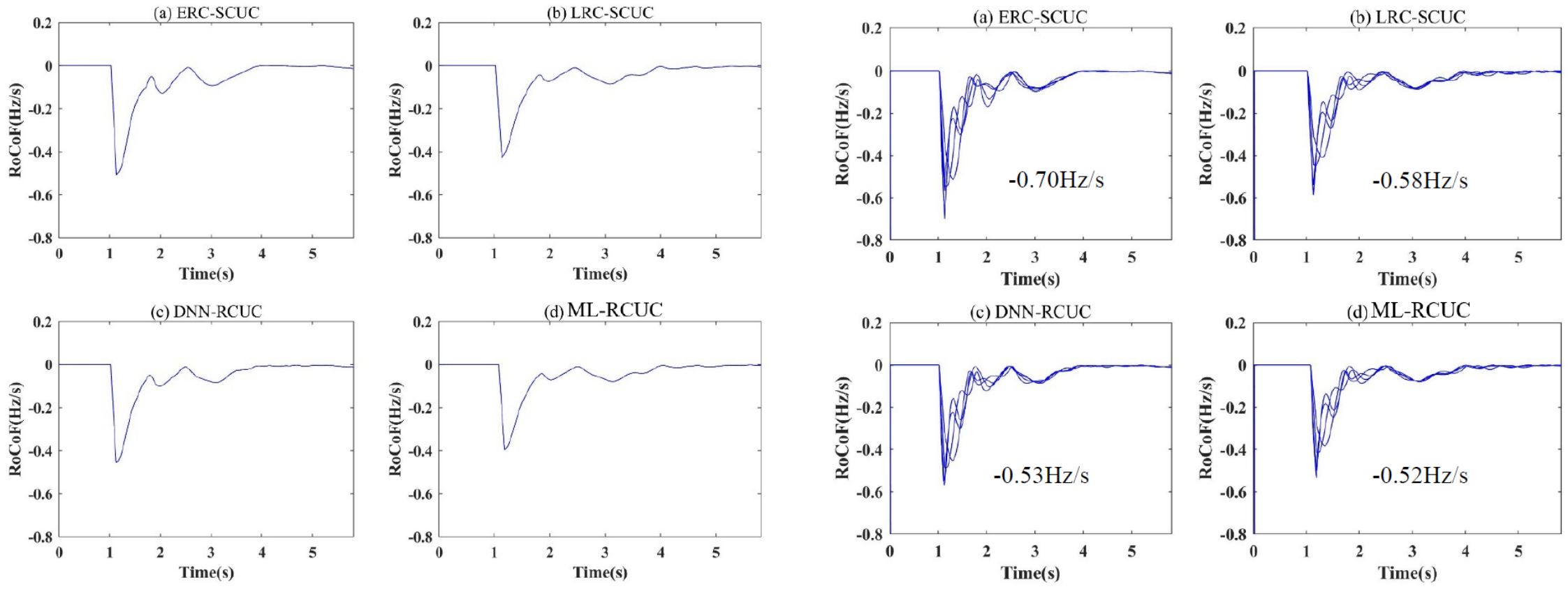

**Fig. 8.** Uniform RoCoF curves of different model following worst contingency case.

**Fig. 9.** Locational RoCoF curves of different model following worst contingency.

## VII. Conclusions and Future Work

### A. Conclusions

The presence of a high penetration level of renewable energy sources in the power grid reduces its overall inertia, which could lead to frequency instability during worst-case $G$-1 contingencies. Frequency-related constraints are desired to be established and included in the energy scheduling models such as SCUC to ensure post-contingency frequency stability. Due to theoretical simplifications inherent in their derivation, current physical model-based approaches face limitations. They either struggle to maintain locational RoCoF stability due to approximation errors or provide overly conservative solutions that result in additional costs. This paper proposes a novel ML-RCUC model that integrates frequency-related constraints derived from convolutional neural networks capturing spatial correlations. Simulations conducted with PSS/E demonstrate that the proposed ML-RCUC model effectively ensures system frequency stability without unnecessary conservatism, showing better performance compared to four other benchmark SCUC models.

### B. Future Work

Correlations exist among load data and generation data, which may affect the performance machine learning-based approaches; in the future, we plan to integrate historical correlation matrices to enhance data realism when generating the training dataset. Key generator control parameters if varying during short-term operations may also affect the performance of the proposed method; a potential future work would include those control parameters in the extended feature set, particularly for scenarios with significant control parameter tuning. While this work focuses on generator contingency, in the future, we will consider transmission line fault and evaluate their impacts. Additional tests on large-scale power system cases would be necessary to demonstrate the scalability of the proposed method. In addition to frequency stability, other types of grid stability are also very important and may also be largely affected by grid revolutions, which could be another potential future research direction.

## ACKNOWLEDGMENT

This research was in part supported by the National Science Foundation (NSF) under Grant No. 2337598; and it was also partly supported by the NSF AI Institute for Advances in Optimization (Award 2112533). Any opinions, findings, and conclusions or recommendations expressed in this paper are those of the author(s) and do not necessarily reflect the views of the NSF.